\documentclass[11pt]{article}

\usepackage{acl}
\usepackage{times}
\usepackage{latexsym}
\usepackage{tikz}
\usetikzlibrary{positioning,arrows.meta}

\usepackage[T1]{fontenc}

\usepackage[utf8]{inputenc}

\usepackage{microtype}

\usepackage{inconsolata}
\usepackage{textgreek}

\usepackage{graphicx}
\usepackage{booktabs}
\usepackage{tabularx}
\usepackage{ragged2e}
\usepackage{amsmath}
\usepackage{amssymb}
\title{Directional Alignment and Narrative Agency in Human–LLM Co-Writing}

\author{
Halfdan Nordahl Fundal \\
Center for Contemporary Cultures of Text \\
Aarhus University \\
\texttt{halfi@cc.au.dk}
\And
Yuri Bizzoni \\
Center for Contemporary Cultures of Text \\
Aarhus University \\
\texttt{yuri.bizzoni@cas.au.dk}
}

\begin{document}
\maketitle
\begin{abstract}
We investigate narrative agency in human–LLM creative co-writing, asking who drives story development in turn-based collaboration. Using a new corpus of 87 human–LLM co-written stories, we apply sentiment and semantic modeling to quantify affective alignment and semantic novelty in turn-taking, and directional measures to assess which agent shapes narrative progression.
Our results show asymmetric influence: human turns introduce greater semantic novelty and are more likely to shape subsequent developments, whereas LLM contributions predominantly elaborate on human-introduced elements. At the sentiment level, alignment is also asymmetric, but more bidirectional: LLMs exhibit stronger turn-level emotional adaptation than humans, but both agents track each other's emotional valence and LLMs show an independent tendency to more positive emotional baselines. 
These findings indicate a complementary division of labor in human–LLM co-writing, where humans drive narrative innovation and direction, while LLMs act as adaptive amplifiers that sustain coherence and elaborate emerging narratives.

\end{abstract}

\section{Introduction}

Large Language Models (LLMs) are becoming frequent ``collaborators" in creative writing practices, supporting both story ideation and stylistic transformation. Prior work has evaluated these systems in terms of output quality and controllability, treating model contributions as isolated outputs and overlooking the \textit{interactive} dynamics that shape a joint narrative.

Creative co-writing is a complex process of negotiation over several dimensions at once, such as emotional tone, content, as well as the story's overall narrative direction. In human--human collaboration, such negotiation involves the necessity of alignment and influence between participants, as well as tensions between different creative aims. As we begin to interact with LLMs as social partners, human--LLM collaboration raises parallel questions: do models tend to adapt to human cues, or do they introduce new trajectories? How does narrative agency work in mixed human--AI teams?

Understanding these dynamics is crucial for both theoretical and practical reasons. From a cognitive and literary perspective, psycholinguistic evaluations of co-writing offer a window into affective coordination and shared meaning between humans and artificial agents. From a design perspective, insights into alignment and influence can inform the development of writing systems that better support creativity or user control. From a digital humanities perspective, studying the affective and informational dynamics of co-authored narratives extends computational approaches to literary sentiment analysis into interactive, multi-agent settings.

In this paper, we introduce a new corpus of co-created human-LLM turn-taking narratives, and examine different aspects of the dynamics that emerge between humans and LLMs through narrative co-creation. Specifically, we assess \textit{emotional alignment} through sentiment modeling between inputs, and evaluate narrative \textit{agency} as the relationship between \textit{novelty}, as information-theoretic deviation of inputs from a baseline, and \textit{resonance}, as the tendency of novel elements to remain active in the subsequent development of the story.  

We address the following research questions:

\begin{itemize}
    \item \textbf{RQ1 (Baseline differences):} Do human and LLM turns differ in mean valence or valence distribution?
    \item \textbf{RQ2 (Affective coordination):} Do human and LLM interlocutors align (turn-to-turn), and is alignment symmetric?
    \item \textbf{RQ3 (Narrative influence):} Whose novel contributions are most likely to be taken up in the next turn - humans' or LLMs'?
\end{itemize}

Our contribution is thus threefold: (1) a new corpus of human–LLM co-written narratives, (2) directional metrics for affective alignment, and (3) novel applications of information-theoretic methods to quantify linguistic influence, supported by empirical evidence of asymmetric narrative agency in mixed human–AI creative collaboration.

\section{Related Work}
Although LLMs continue to show impressive performance across NLP benchmarks, research on their creative output is mixed and often highlights limitations in open-ended tasks, such as generating diverse and dynamically evolving narratives \cite{tian-etal-2024-large-language}. Such studies often investigate isolated LLM output \cite{arora2025generative}, overlooking the iterative, relational nature of most LLM use. 

\paragraph{Human-LLM creativity}
Findings on human--AI joint creativity remain mixed on whether LLMs increase or inhibit creativity. Access to AI tools can increase absolute contribution of novel artifacts due to the productivity effect \cite{zhou2025expands}, and enhance overall writing quality \cite{noy2023experimental} compared to the individual. On the other hand, human-LLM interactions in creative fields can result in creative fixation in complex tasks \cite{cheng2025inspiration} and homogenization of output by narrowing the diversity of ideas \cite{anderson2024homogenization}. Ultimately, human-LLM iteration seems to outperform the individual in creative output, while consistently underperforming human-human collaboration \cite{tang2025best}. Earlier work on human-LLM interaction paradigms has explored facilitation of collaborative co-creative frameworks through dyadic storytelling \cite{roemmele2015creative, clark2021choose}, but did not touch on the emergent dynamics between agents. Although summary output evaluates general differences of joint creativity, it leaves unclear how the dynamics unfold during collaboration, and how collaborative roles emerge. 

\paragraph{Emotional contagion and alignment}
Collaborative processes, such as joint storytelling, involve negotiation of sentiment, agency, and influence. In conversational paradigms, interlocutors mimic and synchronize behavior, often referred to as emotional contagion \cite{hatfield1993emotional}. \citet{varni2017computational} proposed a computational framework to quantify emotional contagion between humans in conversation. They modeled the alignment as matches between interlocutor time series polarity states (positive, negative, neutral), using valence and facial expressions as metrics. \citet{poria2019emotion} highlighted how challenges to model emotion recognition in text can improve conversational AI-systems, and emphasized how interlocutors exert emotional influence both on their counterpart and on themselves. Studies on human-LLM dialogues have shown convergence in speech rate \cite{li2025ai} and linguistic convergence \cite{wilkenfeld2022ai}.  

\paragraph{Sentiment analysis in digital humanities}
Computational sentiment analysis has become a tool for studying emotional dynamics in literary texts. \citet{reagan2016emotional} used a sliding-window computational framework to identify six dominant emotional arcs, while other approaches modeled progression \cite{hu2021dynamic} or mood \cite{ohman2022computational}. In domain-specific settings, \citet{feldkamp2024comparing} compared dictionary-based and transformer-based sentiment tools for Danish literature, while \citet{Bizzoni2023ComparingTA} evaluated similar methods on Hemingway's works, highlighting the challenges of applying general-purpose sentiment models to literary text. \citet{lyngbaek2025continuous} introduced concept vector projection as a method for deriving continuous sentiment scores tailored to literary and multilingual contexts, which we adopt in the present study. Our work extends this line from single-author literary analysis to interactive, multi-authored narrative frameworks.

\paragraph{Influence and agency}
On agency and influence, \citet{barron2018individuals} presented an approach to evaluate the influence of novel ideas on subsequent discourse, introducing the measures of novelty, transience, and resonance. They used a corpus of speeches during the French Revolution, and, through KL-divergence of topic distributions, captured novelty as a speech's deviation from previous discourse, transience as the degree to which that deviation failed to persist, and resonance as the difference between the two, expressing the deviation that is retained in future discourse. \citet{bergey2024yeah} focused on information propagation in dialogue, and used token-level LLM surprisal scores to model conversational information flow between interlocutors. We build on both approaches, applying surprisal-based novelty, transience, and resonance to human--LLM co-writing to evaluate whose contributions exhibit greater narrative influence.

\section{A Corpus of Narrative Co-Writing}

\subsection{Task and experiment design}
\begin{table}[t]
\centering
\small
\setlength{\tabcolsep}{6pt}
\renewcommand{\arraystretch}{1.2}
\begin{tabular}{l r}
\toprule
\textbf{Corpus metric} & \textbf{Value} \\
\midrule
Total stories ($n$) & 87\\
Total participants ($n$) & 87\\
Interactions per story& 10 \\
Mean LLM word count& 29\\
 Mean user word count&26\\
Total interactions ($n$) & 870\\
\bottomrule
\end{tabular}
\caption{Descriptive statistics for the co-writing corpus.}
\label{tab:corpus_descriptives}
\end{table}

\begin{figure}[t]
  \centering
  \includegraphics[width=\linewidth]{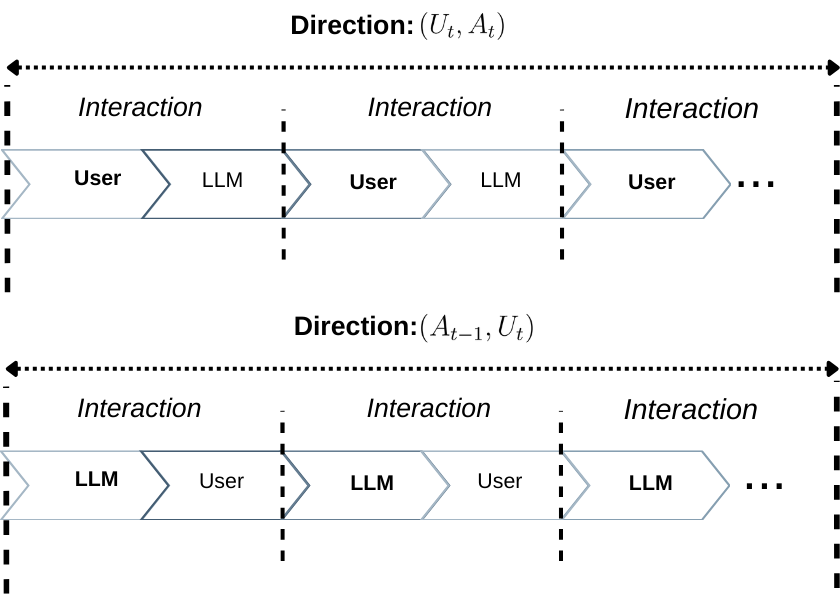}
  \caption{Dyadic task flow, visualizing how the participant and LLM take turns adding to the narrative.}
  \label{fig:task_flow}
\end{figure}

We designed a controlled collaborative storytelling task that isolates turn-based interaction dynamics. The task models co-writing as a dyadic exchange in which a human participant and one of four LLMs jointly construct a narrative through alternating contributions\footnote{Participants were randomly assigned to one of the following LLMs as writing partners: gpt-4.1-2025-04-14, claude-sonnet-4-5-20250929, Llama-3.3-70B-Instruct, or Qwen2.5-72B-Instruct.}.
Each agent, human or LLM, contributed to the story from where the other left off, for a total of 10 interaction steps, each comprising one user turn and one LLM turn ($U_t, A_t$), yielding 20 turns per story.
Figure~\ref{fig:task_flow} illustrates the task flow. 

The resulting dataset is structured into four levels of analysis:

\begin{enumerate}
    \item \textbf{Token level (model level):} individual tokens $w$.
    \item \textbf{Turn level:} a single agent contribution at turn $t$, denoted $I_t$, which can be further specified as either a user turn ($U_t$) or an LLM turn ($A_t$).
    \item \textbf{Interaction level:} paired contributions at the same turn $t$, represented as $(U_t, A_t)$.
    \item \textbf{Story level:} a sequence of 10 interactions for session $i$, denoted $S_i$.
\end{enumerate}

The following instructions were provided to both participants and the LLM:

\begin{quote}
\textit{“You are an author taking part in a collaborative storytelling activity with another author. Together, you will create a story by taking turns adding to it. Your goal is to continue from where your partner has left off. If there is no story, please begin the story. You have 10 interactions to write the story. Your input may be slightly truncated by a random number of characters.”}
\end{quote}

The study comprised a total of 91 participants, with a balanced gender distribution (male = 44, female = 46, other = 1) and an age distribution of $M = 24$ years ($SD = 10.4$). This yielded a corpus of 91 stories (910 interactions).\footnote{The corpus will be made publicly available upon publication.} 
Each story consists of alternating human and model contributions of approximately 25--30 words. Examples of the stories and the metrics of each turn are in Table \ref{tab:example_interaction_features}.

\begin{figure}[t]
  \centering
  \includegraphics[width=\linewidth]{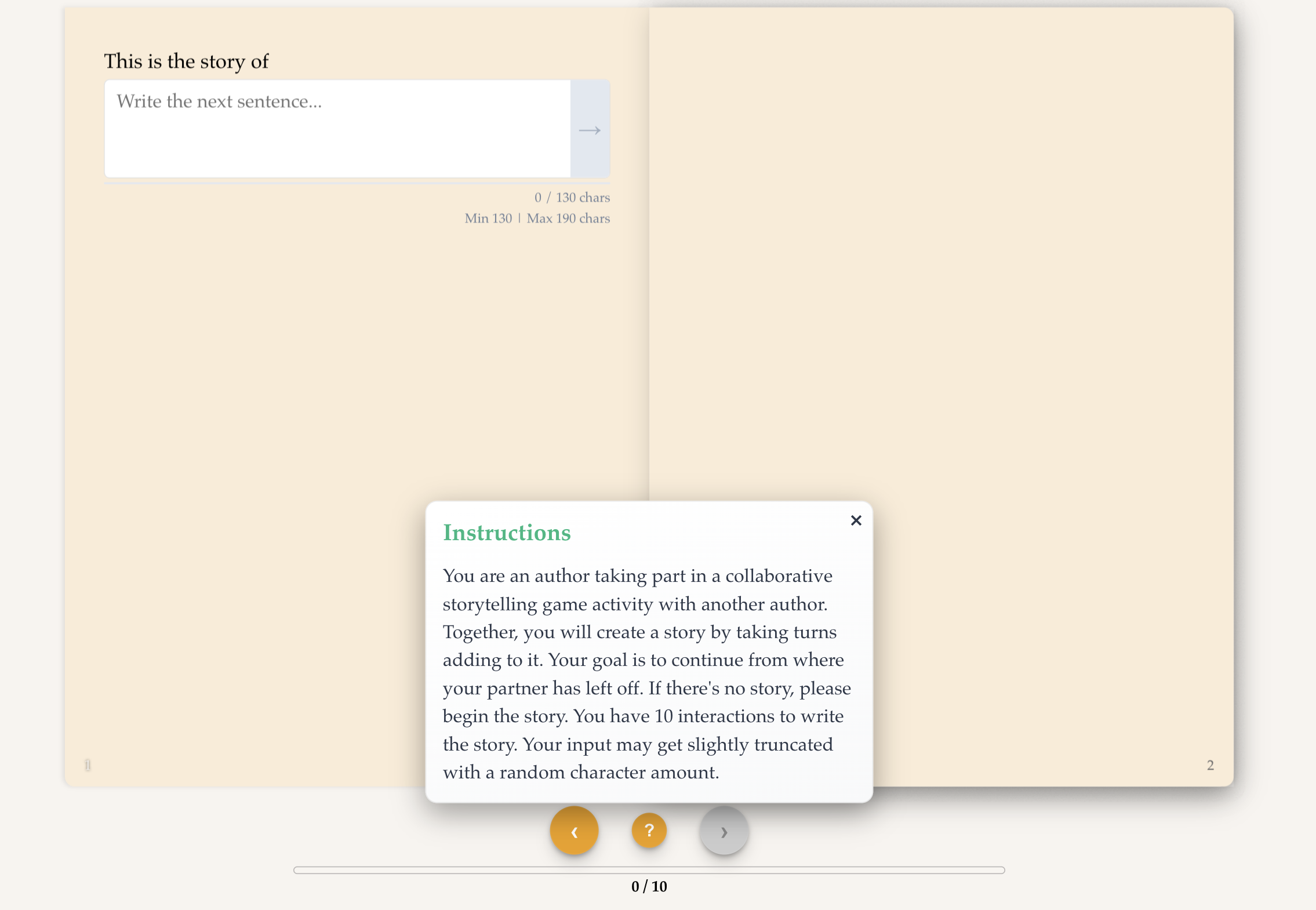}
  \caption{The user interface of the platform used in the experiment with their instructions.}
  \label{fig:ui_screenshot}
\end{figure}

\begin{table*}[t]
\centering
\scriptsize
\setlength{\tabcolsep}{3pt}
\renewcommand{\arraystretch}{1.15}
\begin{tabular}{p{0.22\textwidth} p{0.22\textwidth} r r r r r r}
\toprule
\textbf{User (excerpt)} & \textbf{AI (excerpt)} &
\textbf{User val.} & \textbf{AI val.} &
\textbf{User tok.} & \textbf{AI tok.} &
\textbf{User nov.} & \textbf{AI nov.} \\
\midrule
This is the story of a small-town girl... &
She feels trapped in the mundane routine... &
$-0.241$ & $-0.327$ & 28 & 31 & $-0.547$ & $-1.545$ \\
Fueled by her reading habit, by books like... &
...she's determined to find a way to break free... &
$0.361$ & $0.667$ & 28 & 30 & $-1.389$ & $-2.038$ \\
...a more fulfilling future. Everything from... &
...though she quickly dismisses the latter... &
$0.804$ & $-0.786$ & 25 & 30 & $-1.336$ & $-2.374$ \\
mild-mannered mother, Puritanical father... &
...traveling the world as a missionary's wife... &
$-0.165$ & $0.066$ & 21 & 30 & $-0.376$ & $-3.067$ \\
smuggled out of Pakistan in the hollow tubing... &
...leaving the small town girl to wonder... &
$-0.673$ & $-0.032$ & 24 & 31 & $-1.209$ & $-2.633$ \\
\bottomrule
\end{tabular}
\caption{Example interaction-level features (valence, token counts, novelty) for a single story.}
\label{tab:example_interaction_features}
\end{table*}

\section{Methods}
\subsection{Preprocessing}
To ensure accurate downstream analysis, a cleaning pipeline was applied to all text data. First, incomplete stories, classified as stories with fewer than 10 total interactions, were excluded. User-generated text was then spell-corrected using GPT-4o-mini (temperature = 0) via API, with instructions to preserve each sentence as it was, correcting only spelling errors. LLM-generated text required no cleaning as it contained no spelling errors. To ensure corrections did not substantially alter the original text, the Levenshtein edit distance was calculated between original and corrected versions, and stories with user text exceeding an edit distance of 70 were excluded as these cases corresponded to users writing gibberish rather than coherent text. In total, four stories were removed, yielding a final analytical sample of 87 stories. Manual post-hoc inspection of spell-corrected user text confirmed minimal syntactic and semantic changes were made in the cleaning process.

\subsection{Valence}
First, we investigate the general difference in emotional tone between agents. Let $I_{t}^{(a)}$ denote the textual content of agent $a \in \{\mathrm{User}, \mathrm{LLM}\}$ at turn $t$. To get a continuous metric of valence, we compute a scalar valence score for each turn using concept vector projection. Using a sentence-embedding model (\texttt{paraphrase-multilingual-mpnet-base-v2}), each turn is embedded as a vector representation $\mathbf{e}_{t}^{(a)} \in \mathbb{R}^d$. The model was chosen for its strong performance across varying registers and vocabulary.
We then compute valence as the scalar projection onto a sentiment concept vector $\mathbf{c}$, constructed as the normalized difference between mean embeddings of positive and negative seed words, following \citet{lyngbaek2025continuous}:
\begin{equation}
v_{t}^{(a)} = \frac{\mathbf{e}_{t}^{(a)} \cdot \mathbf{c}}{\|\mathbf{c}\|}.
\end{equation}
Intuitively, this projection measures how far each turn's embedding lies along the positive--negative sentiment axis in the embedding space. Higher values indicate more positively-toned language, while lower values indicate more negative tone. To evaluate baseline differences in emotional tone between agents, we fitted a linear mixed-effects model:
\begin{equation}
v_{it} = \beta_0 + \beta_1 \mathrm{Agent}_{it} + (1 \mid \mathrm{Story}_i),
\end{equation}
where $v_{it}$ denotes the valence score for turn $t$ in story $i$, and $\mathrm{Agent}_{it}$ is a binary indicator distinguishing User from LLM turns. Random intercepts for $Story$ account for the non-independence of turns within stories. This approach to valence estimation allows a stronger domain-fit than standard sentiment analysis methods \cite{lyngbaek2025continuous}, as it can tailor the concept vector to the affective vocabulary most relevant to the data at hand.

\subsection{Directional Sentiment Alignment}
To measure directionality of alignment as an expression of affective coordination between agents, we model the relationship between predictor-turn valence and response-turn valence under two directional pairing rules: 

\begin{enumerate}
    \item \textbf{User$\rightarrow$LLM}: $(U_t, A_t)$
    \item \textbf{LLM$\rightarrow$User }: $(A_{t-1}, U_t)$
\end{enumerate}

Let $v^{(X)}_{it}$ denote the predictor valence and $v^{(Y)}_{it}$ the response valence under a directional pairing rule in story $i$. Directional alignment was estimated using the following linear mixed-effects model:
\begin{equation}
\begin{aligned}
v^{(Y)}_{it} =\;& \beta_0 + \beta_1 v^{(X)}_{it} + \beta_2 D_{it} \\
&+ \beta_3 \left( v^{(X)}_{it} \times D_{it} \right) + (1 \mid \mathrm{Story}_i)
\end{aligned}
\end{equation}
where $D_{it}$ is a binary indicator of alignment direction. Here, $\beta_1$ expresses baseline alignment strength, and the interaction term $\beta_3$ captures directional asymmetry in adaptation of emotional tone.

As two complementary analyses, we (1) computed Fisher $z$-transformed Pearson correlations for each direction and (2) modeled self-alignment as the cosine similarity between an agent's current turn and their own preceding turn, $(U_{t-1}, U_t)$ and $(A_{t-1}, A_t)$.

\subsection{Narrative influence}
\subsubsection{Surprisal}
The surprisal scores for turn-level text are computed using the open-source model Llama-3.1-8B-Instruct as the mean negative log-probability of the tokens in a turn given a preceding context. The single surprisal score of a token is thus calculated as the surprisal score of the token $w_j$ given all preceding tokens in the context window plus the preceding tokens within the turn. 
 \begin{equation}
s(w_j) = -\log_2 p(w_j \mid w_{<j})
\end{equation}
where $j$ indexes tokens within a turn. The mean surprisal score of a turn is then expressed as:
\begin{equation}
\bar{s}(T) = \frac{1}{n} \sum_{j=1}^{n} s(w_j)
\end{equation}

We then define novelty and transience as the difference between conditional and unconditional surprisal. To calculate the pointwise mutual information (PMI) of a turn, we subtract the surprisal score of the turn without context from the surprisal score of the turn with context, resulting in a negative score of contextual facilitation, where more negative values indicate increased predictive benefit from the context. For novelty, the score with context is calculated using all preceding tokens in the story as context, denoted $w_{<t}$, and the surprisal score without context is calculated using the model's beginning-of-sequence token, BOS, providing an unconditional baseline.

\begin{equation}
\mathrm{Novelty}_t = \bar{s}(T_t \mid w_{<t}) - \bar{s}(T_t \mid \text{BOS})
\end{equation}
For computing transience we measure the information gain of the current turn on the full subsequent partner turn, denoted $F_t$. For user turns $F_t$ is the immediate LLM response, and for LLM turns $F_t$ is the next user contribution.
\begin{equation}
\mathrm{Transience}_t = \bar{s}(F_t \mid T_t) - \bar{s}(F_t \mid \text{BOS})
\end{equation}
Finally, resonance was computed by subtracting transience from novelty.
\begin{equation}
\mathrm{Resonance}_t = \mathrm{Novelty}_t - \mathrm{Transience}_t
\end{equation}
\subsubsection{Influence Modeling}
For modelling the relationship between novelty and resonance, we fit a linear mixed-effects model, that evaluates narrative influence by having novelty predict resonance with agent as an interaction effect:
\begin{equation}
\begin{aligned}
\mathrm{Resonance}_{it} ={}& \beta_0 + \beta_1 \mathrm{Novelty}_{it} + \beta_2 \mathrm{Agent}_{it} \\
&+ \beta_3 \mathrm{Novelty}_{it}\mathrm{Agent}_{it}
+ b_{0,\mathrm{story}(i)} \\ &+ \varepsilon_{it}
\end{aligned}
\end{equation}
Supporting this, we also fit a similar mixed effects model, having novelty predict transience with agent as an interaction effect:
\begin{equation}
\begin{aligned}
\mathrm{Transience}_{it} ={}& \beta_0 + \beta_1 \mathrm{Novelty}_{it} + \beta_2 \mathrm{Agent}_{it} \\
&+ \beta_3 \mathrm{Novelty}_{it}\,\mathrm{Agent}_{it}
+ b_{0,\mathrm{story}(i)} \\ &+ \varepsilon_{it}
\end{aligned}
\end{equation}
Because resonance is defined as novelty minus transience, the resonance model is algebraically linked to the transience model. We report both for interpretive transparency, as the transience model isolates the uptake mechanism, while the resonance model captures net narrative influence.
\begin{table*}[t]
\centering
\small
\setlength{\tabcolsep}{5pt}
\renewcommand{\arraystretch}{1.2}
\begin{tabularx}{\textwidth}{l >{\RaggedRight\arraybackslash}p{0.23\textwidth} >{\RaggedRight\arraybackslash}X >{\RaggedRight\arraybackslash}X}
\toprule
\textbf{RQ} & \textbf{Unit of analysis} & \textbf{Measure(s)} & \textbf{Model / Tool} \\
\midrule
RQ1 & Turn $(U_t, A_t)$ &
Valence distributions (sentiment scores per turn) &
\texttt{paraphrase-multilingual-mpnet-base-v2}
 + sentiment concept vector projection 
\\
\midrule
RQ2 &
Directional pairings: $(U_t, A_t)$ and $(A_{t-1}, U_t)$\newline
Self-alignment: $(U_{t-1}, U_t)$, $(A_{t-1}, A_t)$ &
Directional alignment: (i) valence coupling (correlation), (ii) semantic alignment (cosine similarity under pairing rules) &
\texttt{paraphrase-multilingual-mpnet-base-v2} (valence embedding) and \texttt{QZhou-Embedding} (semantic embeddings) \\
\midrule
RQ3 &
Token $w_j$ with conditional context $p(w_j \mid w_{<j})$ (aggregated to turn level) &
Surprisal-based novelty, transience and resonance (informational deviation + uptake) &
\texttt{Llama-3.1-8B-Instruct} (token probabilities $\rightarrow$ surprisal) \\
\bottomrule
\end{tabularx}
\caption{Summary of the employed unit of analysis, measures and models for each research question.}
\label{tab:rq_summary}
\end{table*}

\section{Results}

\subsection{Valence distributions differ (RQ1)}
\begin{figure}[t]
  \centering
  \includegraphics[width=\linewidth]{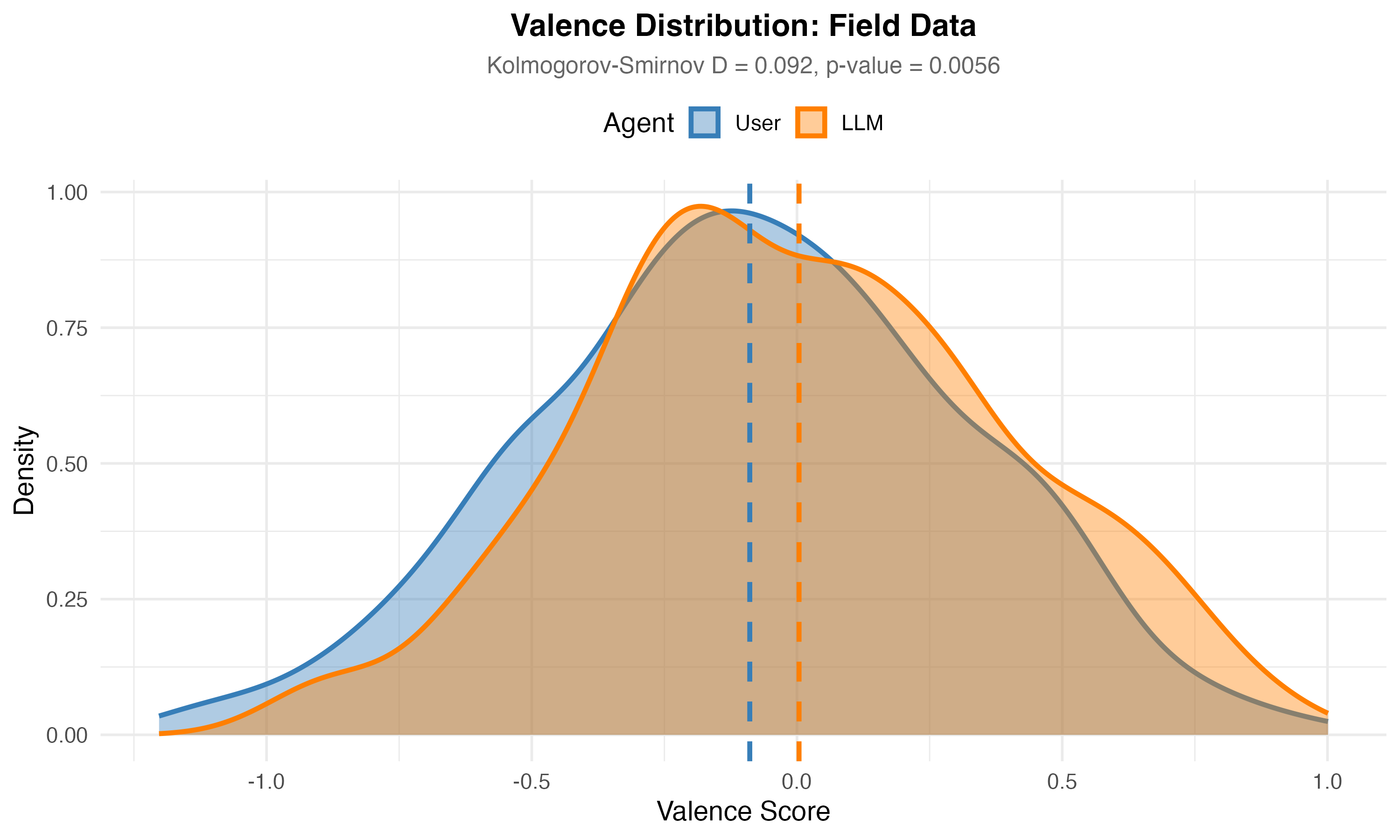}
  \caption{Overlapping density plots showing distributions of valence scores for each agent.}
  \label{fig:valence_dist}
\end{figure}

\begin{figure}[t]
  \centering
  \includegraphics[width=\linewidth]{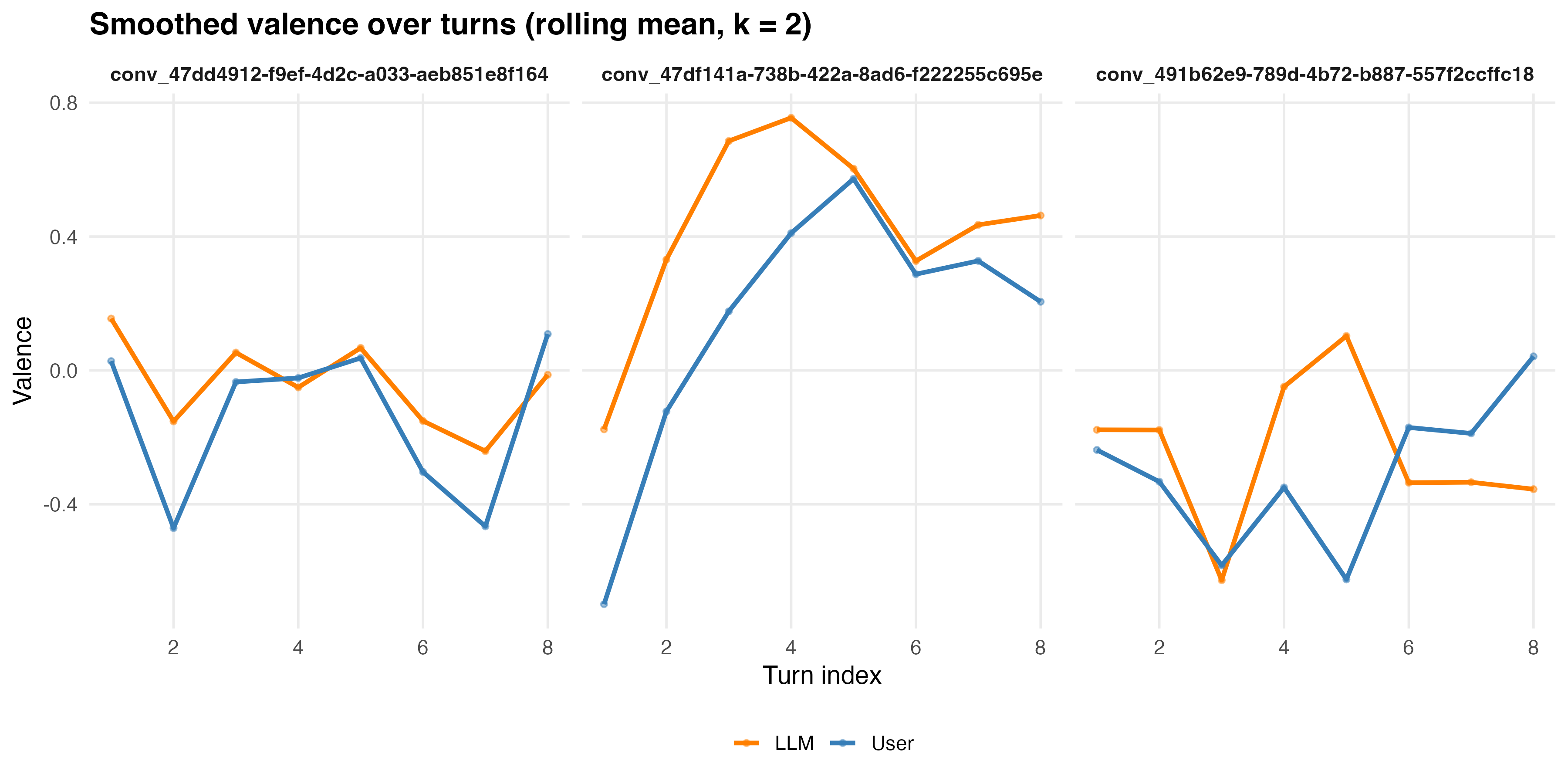}
  \caption{Example of the valence trajectories through three different stories, visualizing baseline differences and temporal alignment.}
  \label{fig:valence_example}
\end{figure}

The distributions of valence showed lower mean user valence ($M=-0.089$, $SD=0.403$) compared to LLM valence ($M=0.004$, $SD=0.397$). A linear mixed-effects model predicting valence from agent type showed a significantly higher mean valence for LLM turns ($\beta = 0.093$, $p<.001$). This coefficient reflects the estimated mean difference between LLM and user turns. Figure~\ref{fig:valence_dist} shows the valence distributions by agent. An example of valence trajectories between agents throughout three stories can be seen in Figure~\ref{fig:valence_example}.

\subsection{Alignment is directional  (RQ2)}
\begin{figure}[t]
  \centering
  \includegraphics[width=\linewidth]{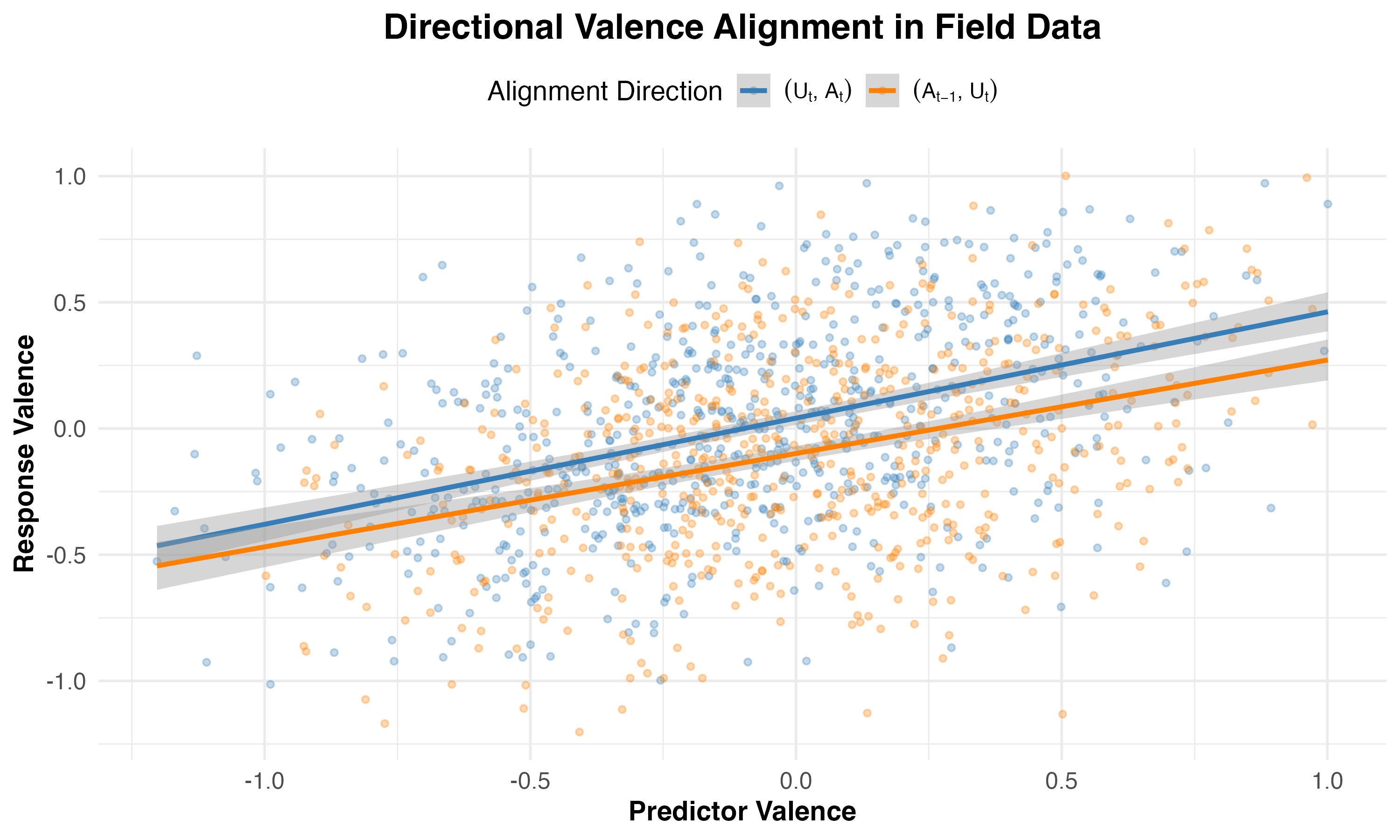}
  \caption{Linear regression of response valence as a function of predictor valence, with color-coded regression slopes for the two directions (blue: $(U_t, A_t)$; orange: $(A_{t-1}, U_t)$).}
  \label{fig:dir_valence}
\end{figure}

The emotional alignment between human and LLM co-writers was statistically significant overall ($\beta = 0.16$, $p < .001$), confirming that valence coordination occurs from turn to turn. However, this alignment was asymmetric. The model showed a strong positive relationship in the $(U_t, A_t)$ direction, with $\beta_1 = 0.232$ ($SE = 0.038$, $p < .001$) compared to the significantly weaker $(A_{t-1}, U_t)$ direction, with a significant interaction contrast ($\Delta\beta=-0.141$, $SE =0.055$, $p= .010$). Figure~\ref{fig:dir_valence} visualizes the fitted relationships by direction.

This was supported by a Pearson correlation analysis. Mean correlation was higher for $(U_t, A_t)$ ($M = 0.232$, $SD = 0.389$) than for $(A_{t-1}, U_t)$ ($M = 0.091$, $SD = 0.395$). One-sample tests against zero indicated that both correlations were reliably above zero, with a stronger effect for $(U_t, A_t)$. For $(U_t, A_t)$: $t = 5.55$, $p <.001$; for $(A_{t-1}, U_t)$: $t = 2.14$, $p = .035$. 

\subsection{Narrative influence is asymmetric (RQ3)}
\begin{figure}[t]
  \centering
  \includegraphics[width=\linewidth]{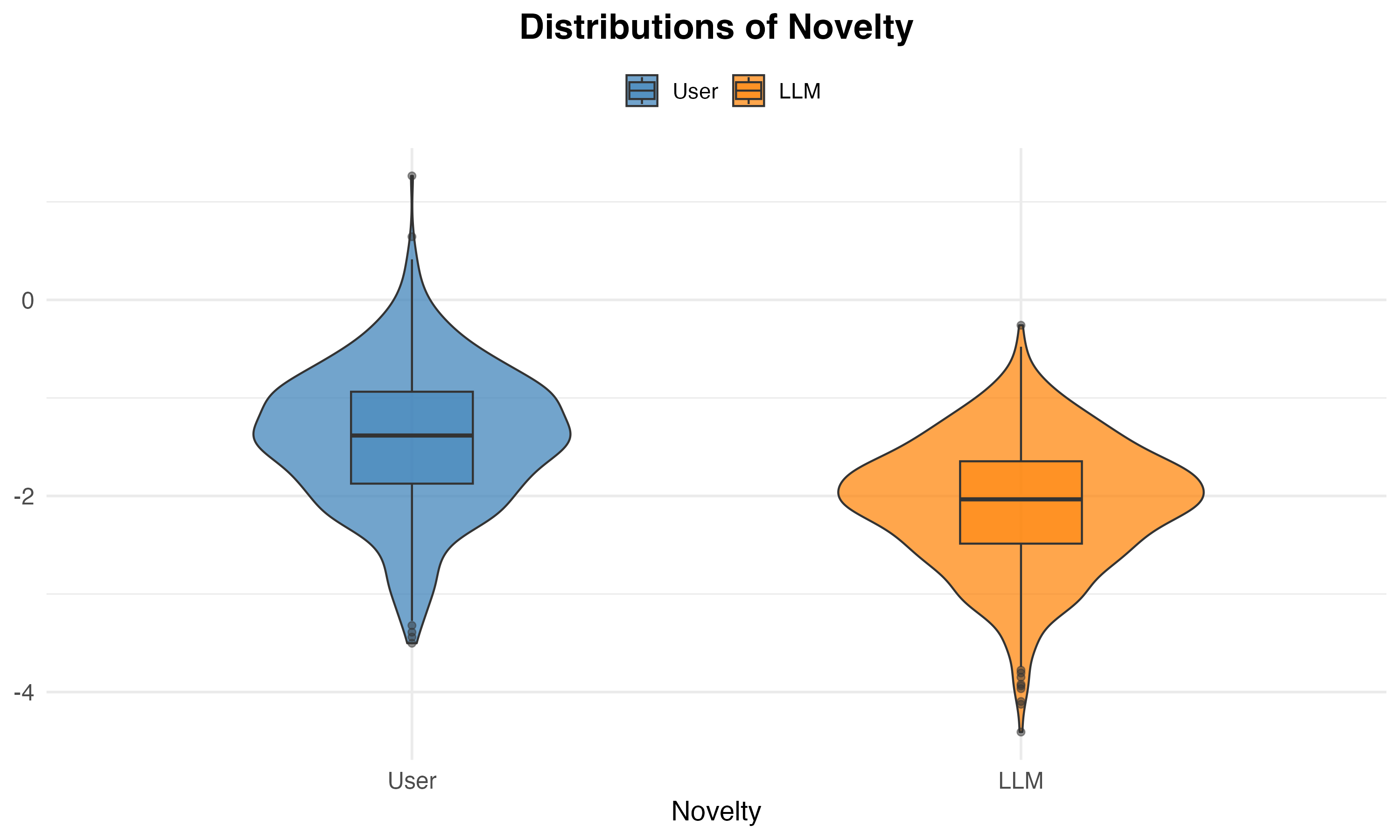}
  \caption{Distributions of surprisal (novelty) for each agent.}
  \label{fig:novelty_dist}
\end{figure}

User turns showed higher novelty than LLM turns (User: $M = -1.418$, $SD = 0.713$; LLM: $M = -2.077$, $SD = 0.647$). User turns also showed slightly higher transience (User: $M = -1.034$, $SD = 0.441$; LLM: $M = -0.676$, $SD = 0.454$). As a result, mean resonance was higher for user turns (User: $M = -0.384$, $SD = 0.774$) than for LLM turns (LLM: $M = -1.401$, $SD = 0.684$). Since novelty is defined as PMI, it is negative when the context reduces surprisal. Figure~\ref{fig:novelty_dist} shows novelty distributions by agent.

\begin{figure}[t]
  \centering
  \includegraphics[width=\linewidth]{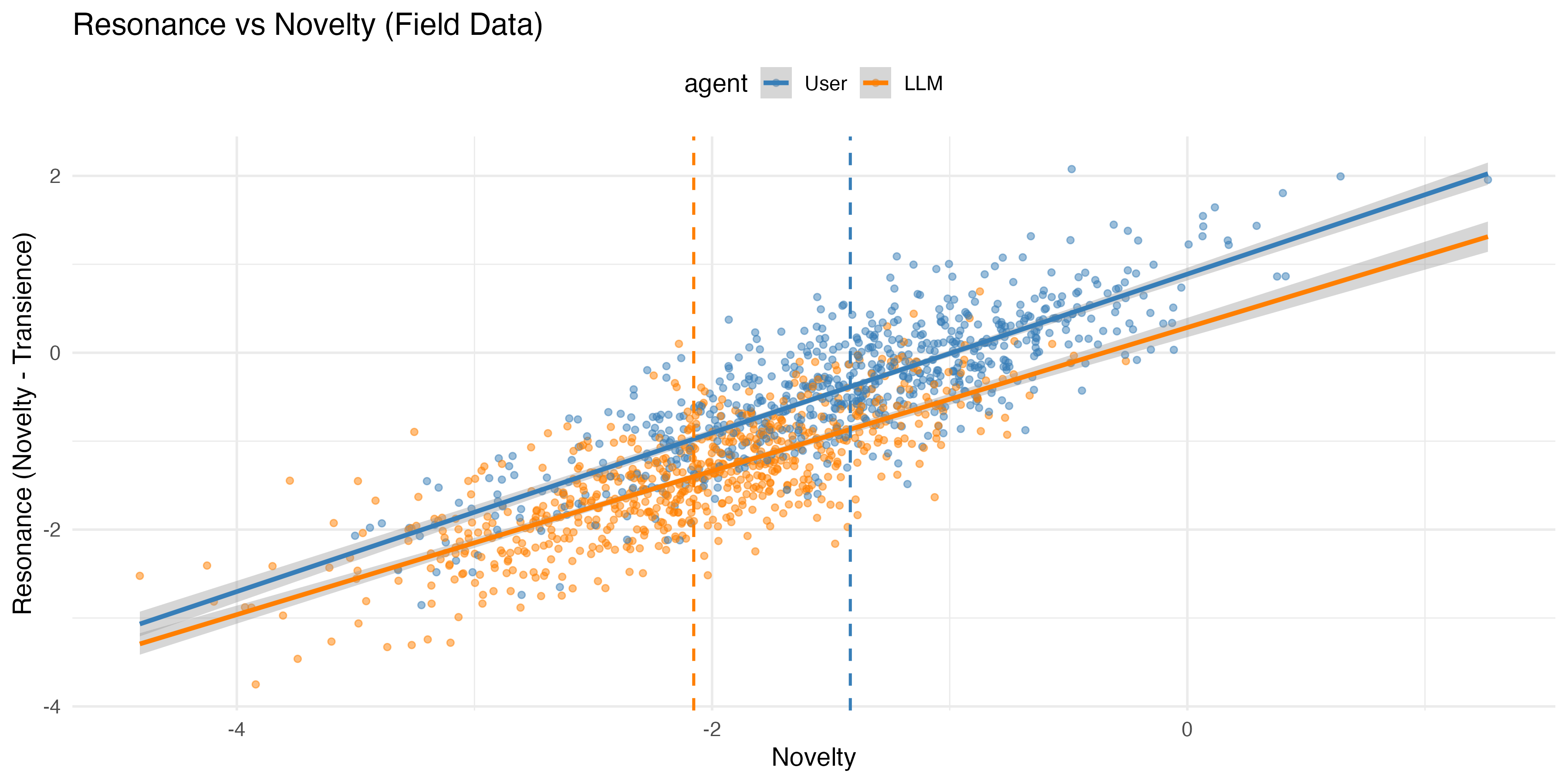}
  \caption{Linear regression of resonance as a function of novelty, with color-coded regression slopes for agent (blue = user, orange = LLM). Dotted lines indicate mean novelty for each agent.}
  \label{fig:resonance_vs_novelty}
\end{figure}

Both agents showed a strong positive novelty-resonance relationship, indicating that more surprising turns tend to stick. In the resonance model, novelty had a strong positive effect for users ($\beta_1 = 0.941$, $p < .001$). The novelty$\times$agent interaction was significant ($\beta_3 = -0.105$, $SE = 0.034$, $p = .0019$), yielding a weaker LLM slope of $0.837$, indicating a slightly weaker innovation bias for LLMs than users. This relationship can be seen in Figure~\ref{fig:resonance_vs_novelty}.

\section{Discussion}
We investigated agent alignment and narrative agency in human--LLM co-writing, and examined affective and semantic adaptation through embedding-based analyses and information propagation through different measures of innovation and influence. Our findings reveal a consistent asymmetry in collaborative dynamics: while both humans and LLMs exhibit alignment and coordination, humans play a disproportionate role in maintaining emotional autonomy, introducing narrative novelty, and shaping subsequent developments.

The emotional asymmetry in valence distributions quantifies a fundamental characteristic of LLMs: their responses tend to drift towards a positive baseline. This could be attributed to the removal of strong negative and profane data during training and fine-tuning of the models. While this has been established for isolated LLM output \cite{munoz2024contrasting,zanotto2024human}, our paradigm tests the autonomy of LLM positivity bias, by having it iteratively react to varying input from human interlocutors, which might force the sentiment of the story into a negative drift. This is equally the case for human partners having to react to positive LLM contributions. This could partly explain the relatively small difference observed in valence distributions, as each partner is faced with a decision to converge emotionally with the counterpart.
\begin{table*}[t]
\centering
\footnotesize
\setlength{\tabcolsep}{10pt}
\renewcommand{\arraystretch}{1.15}
\begin{tabular}{lcccl}
\toprule
\textbf{Result}& 
\textbf{User / $A_{t-1}{\to}U_t$}& \textbf{LLM / $U_t{\to}A_t$}& \textbf{Test} & \textbf{$p$} \\
\midrule
Mean valence & $-0.089$ & $0.004$ & Agent difference (LMM) & $p < .001$ \\
Alignment slope ($\beta$) & $0.091$ & $0.232$ & Slope difference (interaction) & $p = .010$ \\
Alignment frequency (\%) & $49.2$ & $58.9$ & Proportion difference & $p = .001$\\
Mean align. duration (turns) & $1.836$ & $2.189$ & Wilcoxon rank-sum & $p = .039$ \\
\midrule
Resonance $\sim$ novelty slope & $0.941$ & $0.837$ & Slope difference (interaction) & $p = .0019$ \\
Transience $\sim$ novelty slope & $0.059$ & $0.164$ & Slope difference (interaction) & $p = .0019$ \\
\bottomrule
\end{tabular}
\caption{Key results summary. For directional metrics (rows 2--5), the first column reports the $A_{t-1}{\to}U_t$ (LLM-to-User) direction; the second reports $U_t{\to}A_t$ (User-to-LLM). For non-directional metrics (rows 1, 6--7), values are per agent.}
\label{tab:key_results_user_llm_tests}
\end{table*}

The analysis of directional emotional adaptation elaborates on the difference in valence distributions. We found significant bidirectional emotional alignment between co-writers, confirming that human--LLM collaboration involves affective coordination through adaptation of emotional tone. However, the alignment was directional: LLM responses tracked the valence of preceding human turns more strongly than humans tracked LLM affect. This asymmetry indicates that LLMs are more responsive to human emotional cues and suggests that humans maintain a stronger level of emotional autonomy in the co-writing process, resisting immediate emotional adaptation to the LLM counterpart. This is additionally highlighted by the significantly stronger semantic self-alignment of users, in which they converge more with their own preceding input than the LLM responses, adopting less of the semantic profile from the counterpart (see Appendix~\ref{sec:appendix}).

Alignment persistence over time revealed a similar pattern: Alignment occurred more frequently in Human$\rightarrow$LLM transitions, while also retaining longer streaks over time. It seems that LLMs both maintain stronger local affective adaptation, while also sustaining longer-term emotional trajectories (see Appendix~\ref{sec:appendix}).

Across analyses, human contributions were both more novel and yielded higher resonance than LLMs', indicating a greater capacity to introduce narrative elements that persist in the unfolding story. In contrast, LLM contributions tended to elaborate on existing material rather than introduce new semantic directions. This pattern suggests an \textit{initiative asymmetry}, in which humans function as primary drivers of narrative innovation, while LLMs act as adaptive amplifiers that reinforce and extend human-introduced content.\footnote{Importantly, although LLMs exhibited a stronger tendency to follow and elaborate on human contributions, both agents alternated between introducing novelty, transience, and aligning with prior turns, indicating that LLMs are adaptive - rather than passive - participants within collaborative dynamics.}

The positive linear relationship between novelty and resonance characterizes the creative, frictionless nature of the writing task where interlocutors accept and take up the preceding contributions of the counterpart, regardless of the surprising character of this contribution given the prior context. Similarly to the previous findings, this pattern is asymmetric between agents. The effect of very surprising input on future discourse is stronger for users, indicating their writing privilege, where users' novel contributions have more influence on the future discourse, compared to equally novel LLM contributions. Supplementary analyses indicate consistent behavior across the four model subgroups (see Appendix~\ref{sec:appendix}).
These asymmetries seem to mirror a complementary division of labor: humans provide innovation, on which LLMs elaborate, reminiscent of improvisational settings in which one participant introduces motifs and the other stabilizes and develops them.
Finally, the applicability of the directional metrics we introduce extends beyond the presented corpus. These approaches may be further applicable as general tools for modeling interlocutor dynamics in conversation, multi-authored text, or political discourse.

\section{Conclusion and Future Work}
This study examined alignment dynamics and narrative agency in human–LLM collaborative storytelling. Our findings reveal clear patterns that characterize human–LLM co-writing: LLMs \textbf{mirror} the emotional tone established by human contributors, \textbf{but shift} it toward a more positive baseline. At the same time, LLMs tend to \textbf{elaborate} on human-introduced ideas \textbf{rather than introduce} novel narrative elements, and have less influence on the subsequent narrative.

In other words, they overall follow the semantics introduced by human writers, but tend to bring a higher valence ``bias" to the story. This behavior suggests that LLMs function effectively as ``following'' collaborators that sustain coherence and reinforce emerging storylines.
Human participants exhibited greater semantic novelty and stronger influence on subsequent narrative developments, indicating a primary role in introducing new directions and maintaining emotional autonomy. Together, these results point to a complementary division of labor: humans drive innovation and narrative, and LLMs provide adaptive elaboration and strong affective alignment.
These findings have implications for the design of human–AI co-creative systems. Interfaces and interaction paradigms may benefit from supporting human control over narrative direction while using LLM strengths in elaboration, stylistic variation, and coherence maintenance. Understanding this complementary dynamic can inform the development of tools that enhance creativity without diminishing human agency.

Future work should extend this paradigm by incorporating human–human baselines to better contextualize mixed collaboration dynamics; exploring longer interaction sequences to capture extended narrative evolution; integrating context-aware sentiment models to account for broader emotional trajectories; and modeling differences between the four models. Investigating how different model architectures, prompting strategies, and genre constraints affect alignment and agency may further clarify the conditions that make human–LLM collaboration most effective.

\section*{Limitations}
Several limitations should be considered when interpreting the results. First, sentiment scores are computed at the turn level and do not incorporate broader narrative context, potentially overlooking longer-range emotional dynamics. Second, surprisal is estimated with respect to the language model used in the analysis and may not fully reflect human perceptions of novelty. Third, the study does not include a human–human co-writing condition, which limits direct comparison between mixed and purely human collaborations. Future work should include this as a baseline to separate LLM-specific effects from general collaborative dynamics.
Additionally, the experimental setup introduces asymmetries in both interaction experience and language processing capabilities between human participants and LLMs, which may influence the observed alignment patterns. Finally, the preprocessing pipeline, including spell correction, may introduce a degree of normalization toward LLM-like text.

\section*{Ethics Statement}
This study was conducted according to ethical guidelines. All participants provided their consent prior to participation and were informed that their text input would be used for research purposes. Only anonymized demographic data (age, gender) were collected. Participants were free to withdraw at any time. Participants were aware that their co-writer was an LLM. The dataset consists of English-language creative fiction produced by adult participants recruited from a university context in Denmark.

\bibliography{custom}

@article{munoz2024contrasting,
  title={Contrasting linguistic patterns in human and LLM-generated news text},
  author={Mu{\~n}oz-Ortiz, Alberto and G{\'o}mez-Rodr{\'\i}guez, Carlos and Vilares, David},
  journal={Artificial Intelligence Review},
  volume={57},
  number={10},
  pages={265},
  year={2024},
  publisher={Springer}
}

@article{zanotto2024human,
  title={Human variability vs. machine consistency: A linguistic analysis of texts generated by humans and large language models},
  author={Zanotto, Sergio E and Aroyehun, Segun},
  journal={arXiv preprint arXiv:2412.03025},
  year={2024}
}

@inproceedings{tian-etal-2024-large-language,
    title = "Are Large Language Models Capable of Generating Human-Level Narratives?",
    author = "Tian, Yufei  and
      Huang, Tenghao  and
      Liu, Miri  and
      Jiang, Derek  and
      Spangher, Alexander  and
      Chen, Muhao  and
      May, Jonathan  and
      Peng, Nanyun",
    editor = "Al-Onaizan, Yaser  and
      Bansal, Mohit  and
      Chen, Yun-Nung",
    booktitle = "Proceedings of the 2024 Conference on Empirical Methods in Natural Language Processing",
    month = nov,
    year = "2024",
    address = "Miami, Florida, USA",
    publisher = "Association for Computational Linguistics",
    url = "https://aclanthology.org/2024.emnlp-main.978/",
    doi = "10.18653/v1/2024.emnlp-main.978",
    pages = "17659--17681"
}

@article{cheng2025inspiration,
  title={Inspiration booster or creative fixation? The dual mechanisms of LLMs in shaping individual creativity in tasks of different complexity},
  author={Cheng, Xusen and Zhang, Lulu},
  journal={Humanities and Social Sciences Communications},
  volume={12},
  number={1},
  pages={1--10},
  year={2025},
  publisher={Palgrave}
}

@article{barron2018individuals,
  title={Individuals, institutions, and innovation in the debates of the French Revolution},
  author={Barron, Alexander TJ and Huang, Jenny and Spang, Rebecca L and DeDeo, Simon},
  journal={Proceedings of the National Academy of Sciences},
  volume={115},
  number={18},
  pages={4607--4612},
  year={2018},
  publisher={National Academy of Sciences}
}

@article{bergey2024yeah,
  title={From" um" to" yeah": Producing, predicting, and regulating information flow in human conversation},
  author={Bergey, Claire Augusta and DeDeo, Simon},
  journal={arXiv preprint arXiv:2403.08890},
  year={2024}
}

@article{tang2025best,
  title={“Who” Is the Best Creative Thinking Partner? An Experimental Investigation of Human--Human, Human--Internet, and Human--AI Co-Creation},
  author={Tang, Min and Hofreiter, Sebastian and Werner, Christian H and Zieli{\'n}ska, Aleksandra and Karwowski, Maciej},
  journal={The Journal of Creative Behavior},
  volume={59},
  number={3},
  pages={e1519},
  year={2025},
  publisher={Wiley Online Library}
}

@inproceedings{anderson2024homogenization,
  title={Homogenization effects of large language models on human creative ideation},
  author={Anderson, Barrett R and Shah, Jash Hemant and Kreminski, Max},
  booktitle={Proceedings of the 16th conference on creativity \& cognition},
  pages={413--425},
  year={2024}
}

@article{noy2023experimental,
  title={Experimental evidence on the productivity effects of generative artificial intelligence},
  author={Noy, Shakked and Zhang, Whitney},
  journal={Science},
  volume={381},
  number={6654},
  pages={187--192},
  year={2023},
  publisher={American Association for the Advancement of Science}
}

@article{arora2025generative,
  title={Generative artificial intelligence models outperform students on divergent and convergent thinking assessments},
  author={Arora, Vikram and Thabane, Alex and Parpia, Sameer and Calic, Goran and Bhandari, Mohit},
  journal={Scientific Reports},
  volume={15},
  number={1},
  pages={36987},
  year={2025},
  publisher={Nature Publishing Group UK London}
}

@article{zhou2025expands,
  title={Who expands the human creative frontier with generative AI: Hive minds or masterminds?},
  author={Zhou, Eric B and Lee, Dokyun and Gu, Bin},
  journal={Science Advances},
  volume={11},
  number={36},
  pages={eadu5800},
  year={2025},
  publisher={American Association for the Advancement of Science}
}

@inproceedings{Bizzoni2023ComparingTA,
  title={Comparing Transformer and Dictionary-based Sentiment Models for Literary Texts: Hemingway as a Case-study},
  author={Yuri Bizzoni and Pascale Feldkamp},
  booktitle={NLP4DH},
  year={2023},
  url={https://api.semanticscholar.org/CorpusID:267410950}
}

@article{varni2017computational,
  title={Computational study of primitive emotional contagion in dyadic interactions},
  author={Varni, Giovanna and Hupont, Isabelle and Clavel, Chloe and Chetouani, Mohamed},
  journal={IEEE Transactions on Affective Computing},
  volume={11},
  number={2},
  pages={258--271},
  year={2017},
  publisher={IEEE}
}

@article{lyngbaek2025continuous,
  title={Continuous sentiment scores for literary and multilingual contexts},
  author={Lyngbaek, Laurits and Feldkamp, Pascale and Bizzoni, Yuri and Nielbo, Kristoffer and Enevoldsen, Kenneth},
  journal={arXiv preprint arXiv:2508.14620},
  year={2025}
}

@article{poria2019emotion,
  title={Emotion recognition in conversation: Research challenges, datasets, and recent advances},
  author={Poria, Soujanya and Majumder, Navonil and Mihalcea, Rada and Hovy, Eduard},
  journal={IEEE access},
  volume={7},
  pages={100943--100953},
  year={2019},
  publisher={IEEE}
}

@article{hatfield1993emotional,
  title={Emotional contagion},
  author={Hatfield, Elaine and Cacioppo, John T and Rapson, Richard L},
  journal={Current directions in psychological science},
  volume={2},
  number={3},
  pages={96--100},
  year={1993},
  publisher={Sage Publications Sage CA: Los Angeles, CA}
}

@inproceedings{feldkamp2024comparing,
  title={Comparing tools for sentiment analysis of Danish literature from hymns to fairy tales: Low-resource language and domain challenges},
  author={Feldkamp, Pascale and Kostkan, Jan and Overgaard, Ea and Jacobsen, Mia and Bizzoni, Yuri},
  booktitle={Proceedings of the 14th Workshop on Computational Approaches to Subjectivity, Sentiment, \& Social Media Analysis},
  pages={186--199},
  year={2024}
}

@article{reagan2016emotional,
  title={The emotional arcs of stories are dominated by six basic shapes},
  author={Reagan, Andrew J and Mitchell, Lewis and Kiley, Dilan and Danforth, Christopher M and Dodds, Peter Sheridan},
  journal={EPJ data science},
  volume={5},
  number={1},
  pages={31},
  year={2016},
  publisher={Springer}
}

@article{hu2021dynamic,
  title={Dynamic evolution of sentiments in Never Let Me Go: Insights from multifractal theory and its implications for literary analysis},
  author={Hu, Qiyue and Liu, Bin and Thomsen, Mads Rosendahl and Gao, Jianbo and Nielbo, Kristoffer L},
  journal={Digital Scholarship in the Humanities},
  volume={36},
  number={2},
  pages={322--332},
  year={2021},
  publisher={Oxford University Press}
}

@inproceedings{ohman2022computational,
  title={Computational exploration of the origin of mood in literary texts},
  author={{\"O}hman, Emily and Rossi, Riikka H},
  booktitle={Proceedings of the 2nd International Workshop on Natural Language Processing for Digital Humanities},
  pages={8--14},
  year={2022}
}

@inproceedings{clark2021choose,
  title={Choose your own adventure: Paired suggestions in collaborative writing for evaluating story generation models},
  author={Clark, Elizabeth and Smith, Noah A},
  booktitle={Proceedings of the 2021 Conference of the North American Chapter of the Association for Computational Linguistics: Human Language Technologies},
  pages={3566--3575},
  year={2021}
}

@inproceedings{roemmele2015creative,
  title={Creative help: A story writing assistant},
  author={Roemmele, Melissa and Gordon, Andrew S},
  booktitle={International Conference on Interactive Digital Storytelling},
  pages={81--92},
  year={2015},
  organization={Springer}
}

@inproceedings{li2025ai,
  title={When AI Speaks, Do We Follow? Phonetic Entrainment in Human-AI Dialogues},
  author={Li, Qixin and Lu, Haocheng and Wang, Gaowu},
  booktitle={National Conference on Man-Machine Speech Communication},
  pages={167--186},
  year={2025},
  organization={Springer}
}

@inproceedings{wilkenfeld2022ai,
  title={“ai love you”: Linguistic convergence in human-chatbot relationship development},
  author={Wilkenfeld, J Nan and Yan, Bei and Huang, Jujun and Luo, Guirong and Algas, Kristina},
  booktitle={Academy of Management Proceedings},
  volume={2022},
  number={1},
  pages={17063},
  year={2022},
  organization={Academy of Management Briarcliff Manor, NY 10510}
}

\appendix

\section{Supplementary Analyses}
\label{sec:appendix}

\subsection{Duration of alignment}
\begin{figure}[t]
  \centering
  \includegraphics[width=\linewidth]{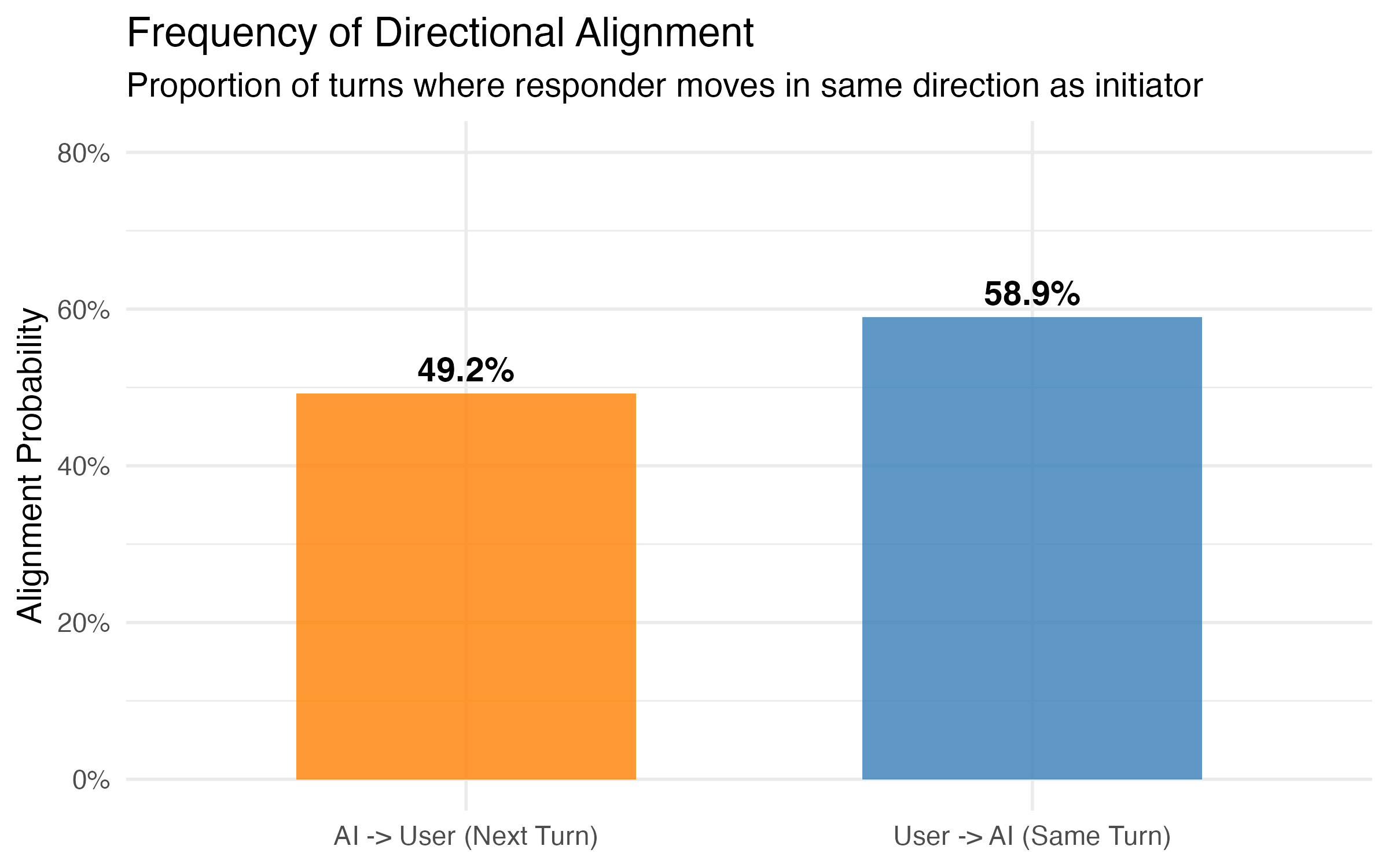}
  \caption{Frequency of directional alignment, shown as the proportion of turns where the responder moves in the same direction as the initiator.}
  \label{fig:align_freq}
\end{figure}

To assess stability of directional alignment, we computed (1) the overall frequency of turns classified as aligned per direction and (2) the duration of consecutive alignment streaks. Figure~\ref{fig:align_freq} summarizes alignment frequency by direction. Alignment frequency differed between directions: alignment occurred on $49.2\%$ of turns in the LLM$\rightarrow$User direction and $58.9\%$ of turns in the User$\rightarrow$LLM direction (two-sample test of proportions, $p=0.001324$).

\begin{figure}[t]
  \centering
  \includegraphics[width=\linewidth]{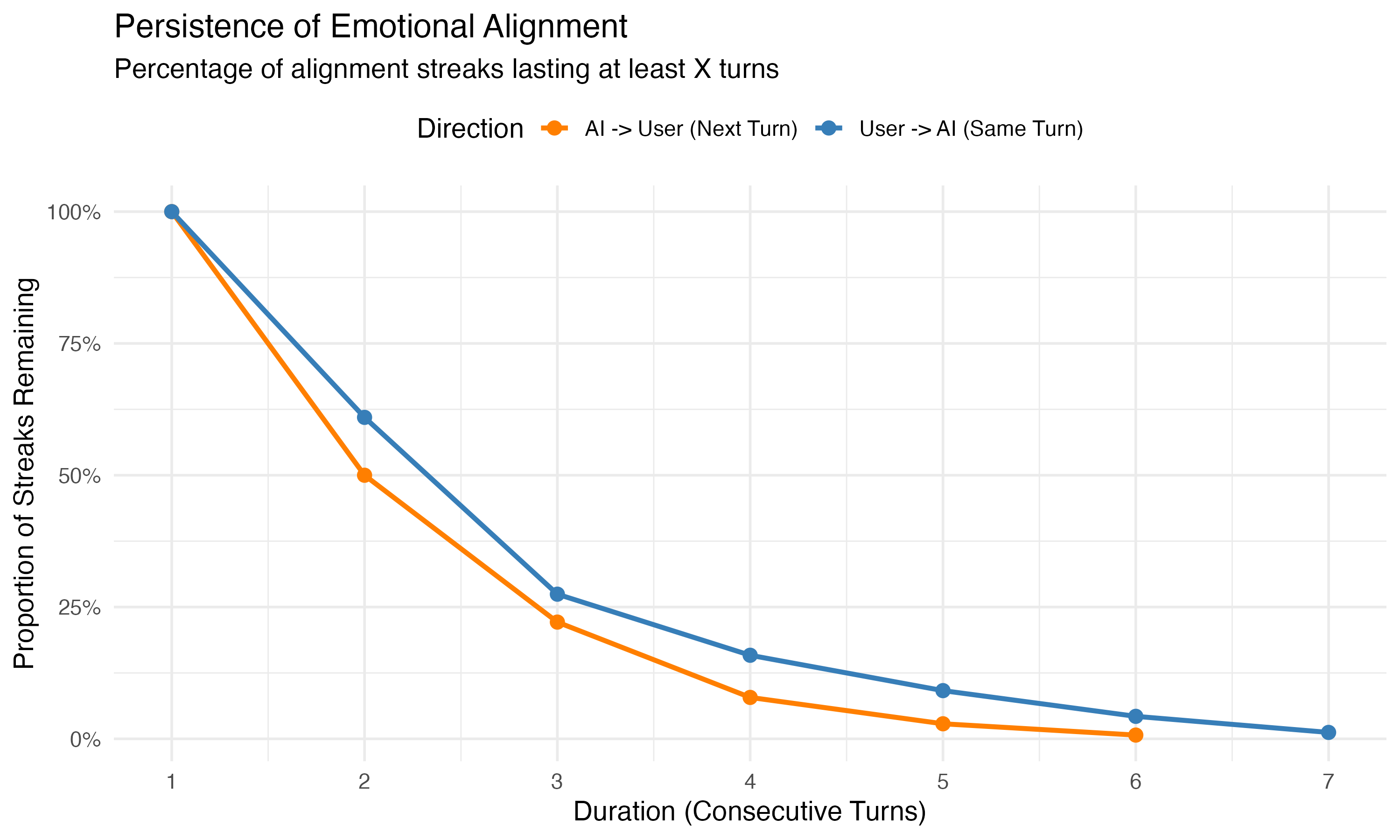}
  \caption{Persistence of emotional alignment. The curves show the percentage of alignment streaks lasting at least $X$ consecutive turns, by direction.}
  \label{fig:align_duration}
\end{figure}

Figure~\ref{fig:align_duration} shows the persistence of alignment streaks (survival-style retention curves). Streaks were longer in the User$\rightarrow$LLM direction (mean $=2.19$, median $=2$, max $=7$) than in the LLM$\rightarrow$User direction (mean $=1.84$, median $=1.5$, max $=6$). A Wilcoxon rank-sum test indicated a reliable difference in duration distributions ($p=0.03943$).

\subsection{Self alignment}
For semantic self-alignment we compared distributions of cosine similarity between $(U_{t-1}, U_t)$  denoting semantic self-alignment of users and $(U_t, A_{t-1})$ denoting semantic alignment of users with the preceding LLM turn. This was similarly modeled for LLM self-alignment. User self-alignment $(U_{t-1}, U_t)$ showed generally higher cosine similarity than user interlocutor alignment $(U_t, A_{t-1})$: the mixed-effects model showed a significant difference ($\beta = -0.012$, $SE = 0.004$, $p = .005$), indicating that users align more with their own preceding turn than with the preceding LLM turn. For LLM self-alignment, no significant difference was observed ($\beta = 0.005$, $SE = 0.004$, $p = .131$).
\begin{figure}[t]
  \centering
  \includegraphics[width=\linewidth]{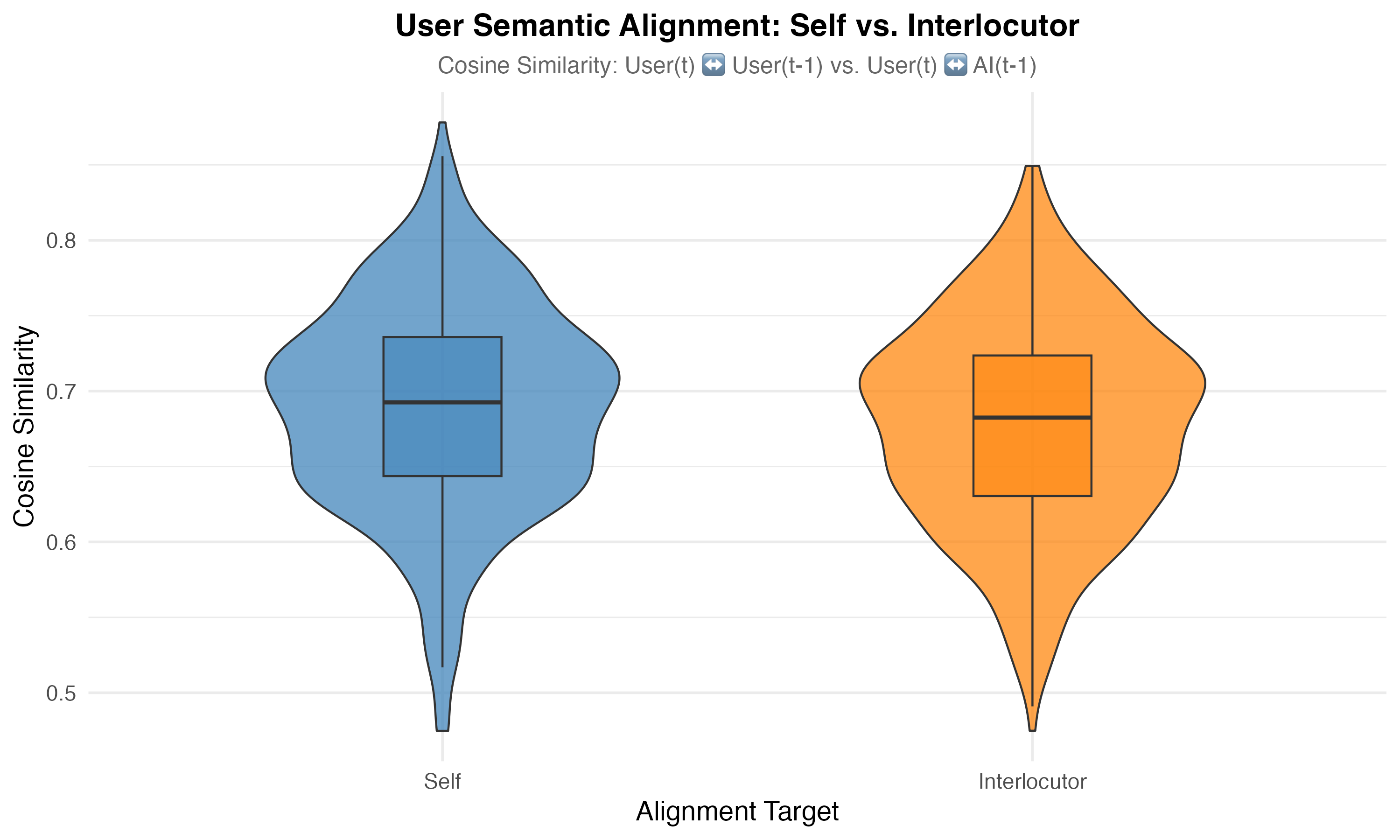}
  \caption{Semantic self-alignment of users (blue) compared to alignment with the interlocutor (orange). Measured as cosine similarity between contributions.}
  \label{fig:align_self}
\end{figure}

\subsection{Alignment and influence by LLM-types}
\begin{figure}[t]
  \centering
  \includegraphics[width=\linewidth]{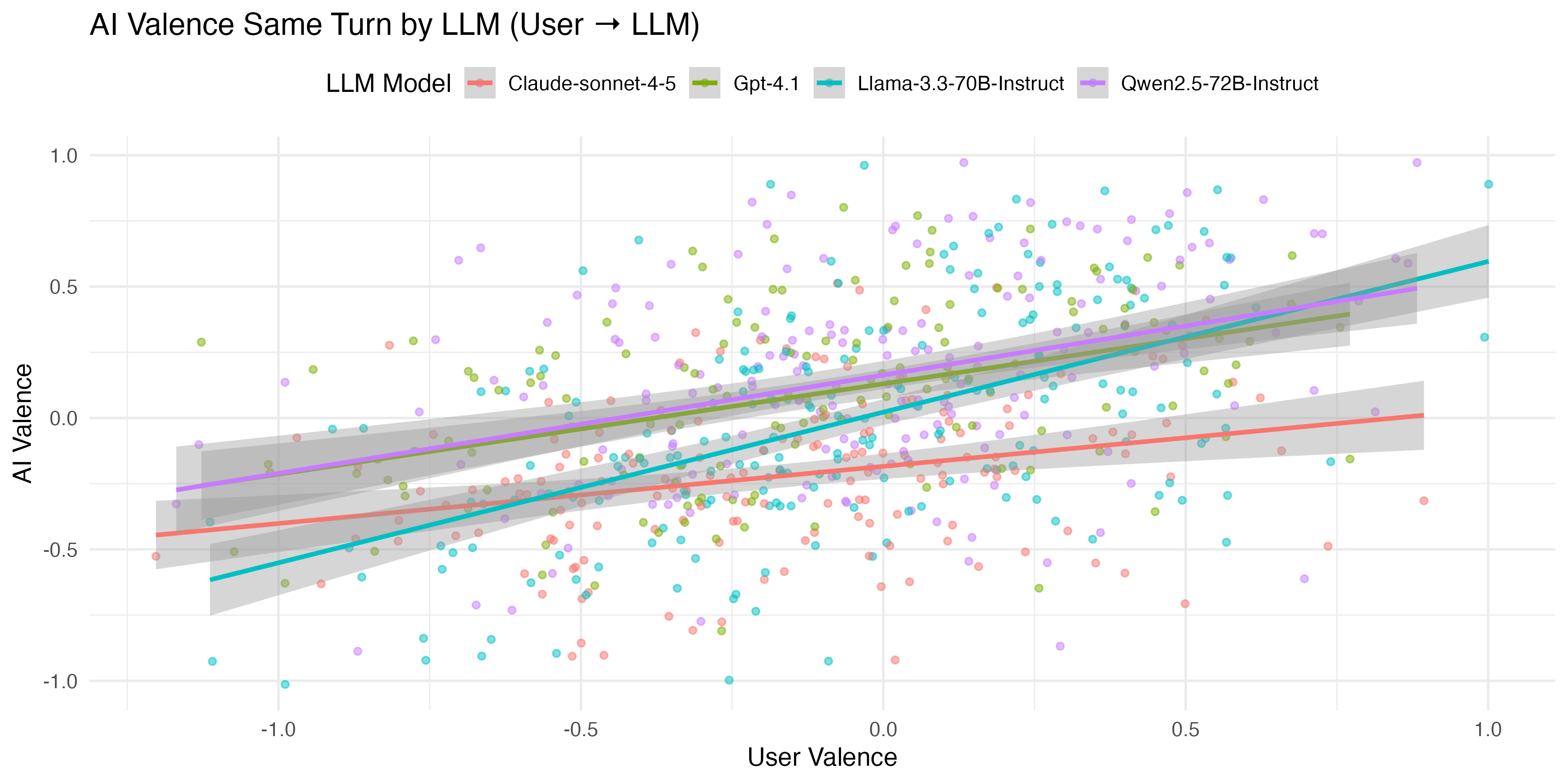}
  \caption{Directional alignment between $(U_t, A_t)$ focusing on LLM alignment, with regression slopes by LLM type.}
  \label{fig:align_llmtype_same}
\end{figure}

\begin{figure}[t]
  \centering
  \includegraphics[width=\linewidth]{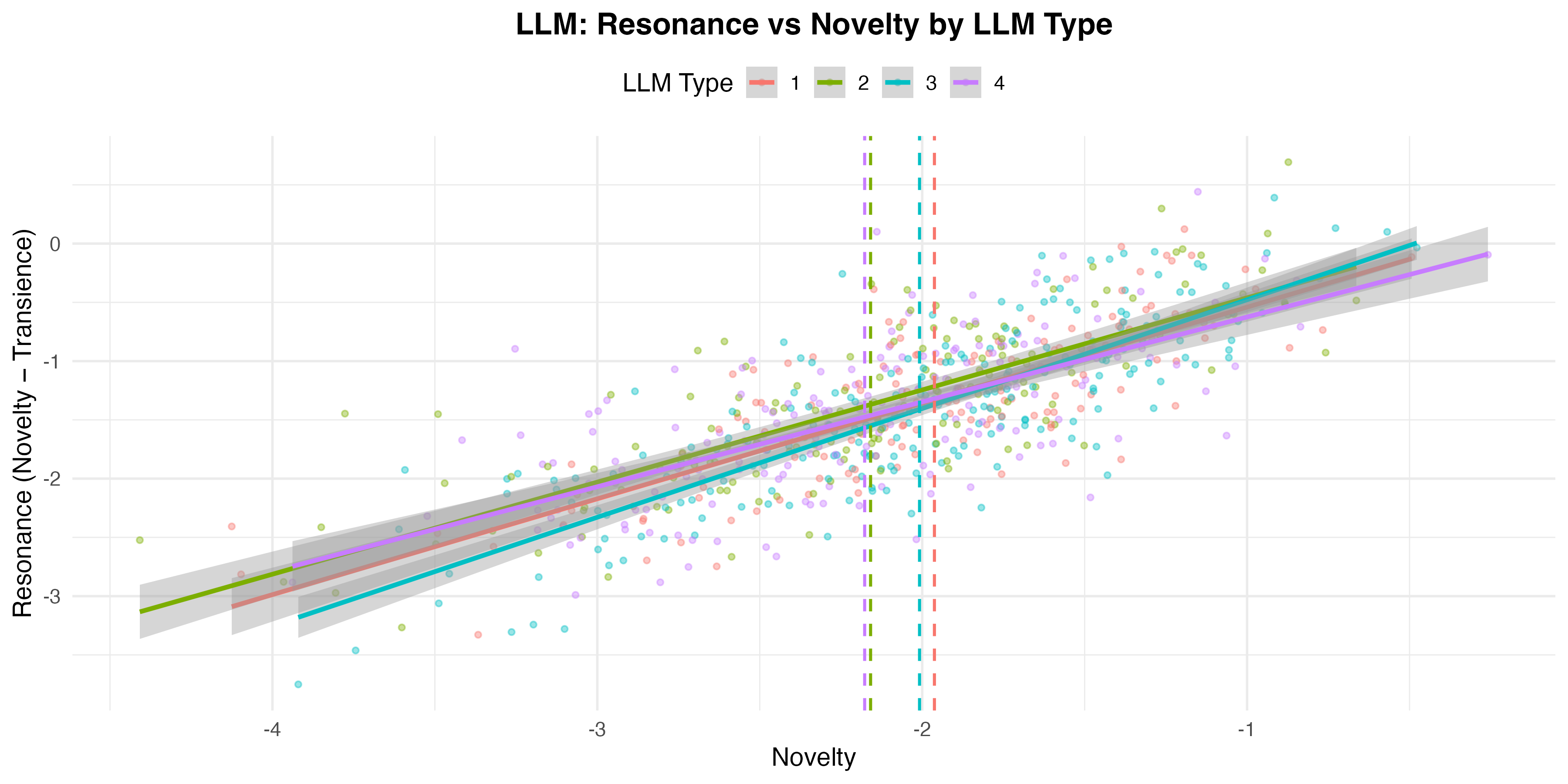}
  \caption{Resonance as a function of novelty for LLM contributions, with color-coded regression slopes for each LLM-type. Dotted lines indicate mean novelty for each model.}
  \label{fig:novelty_llm_type}
\end{figure}

To validate the aggregation of the four different LLMs in the analysis, we conducted inter-model analysis on both alignment $(U_t, A_t)$ and LLM novelty-resonance models. In the analysis of alignment, models showed consistent patterns of sentiment alignment, but with Claude exhibiting a significantly lower intercept than the reference model ($\beta = -0.315$, $SE = 0.062$, $p < .001$), indicating lower mean AI valence in the User$\rightarrow$LLM direction, as seen in Figure~\ref{fig:align_llmtype_same}. Alignment slopes did not significantly differ across models. A one-way ANOVA on per-story same-turn correlations confirmed no significant effect of LLM type ($F(3,83) = 0.40$, $p = .754$). In the novelty-resonance analysis shown in Figure~\ref{fig:novelty_llm_type}, all models showed similar positive novelty-resonance relationships.

\begin{table}[t]
\centering
\footnotesize
\setlength{\tabcolsep}{4pt}
\renewcommand{\arraystretch}{1.15}
\begin{tabular}{lrrrr}
\toprule
\textbf{LLM type} & \textbf{$n$} & \textbf{Mean $\rho$} & \textbf{SD} & \textbf{Mean AI val.} \\
\midrule
GPT-4.1    & 18 & $0.301$ & $0.353$ & $0.085$ \\
Claude     & 20 & $0.198$ & $0.364$ & $-0.219$ \\
Llama 3.3  & 26 & $0.255$ & $0.405$ & $-0.015$ \\
Qwen 2.5   & 23 & $0.180$ & $0.432$ & $0.154$ \\
\bottomrule
\end{tabular}
\caption{Per-story same-turn alignment ($\rho$) and mean AI valence by LLM type. ANOVA: $F(3,83) = 0.40$, $p = .754$.}
\label{tab:llm_type_metrics}
\end{table}

\begin{figure}[t]
  \centering
  \includegraphics[width=\linewidth]{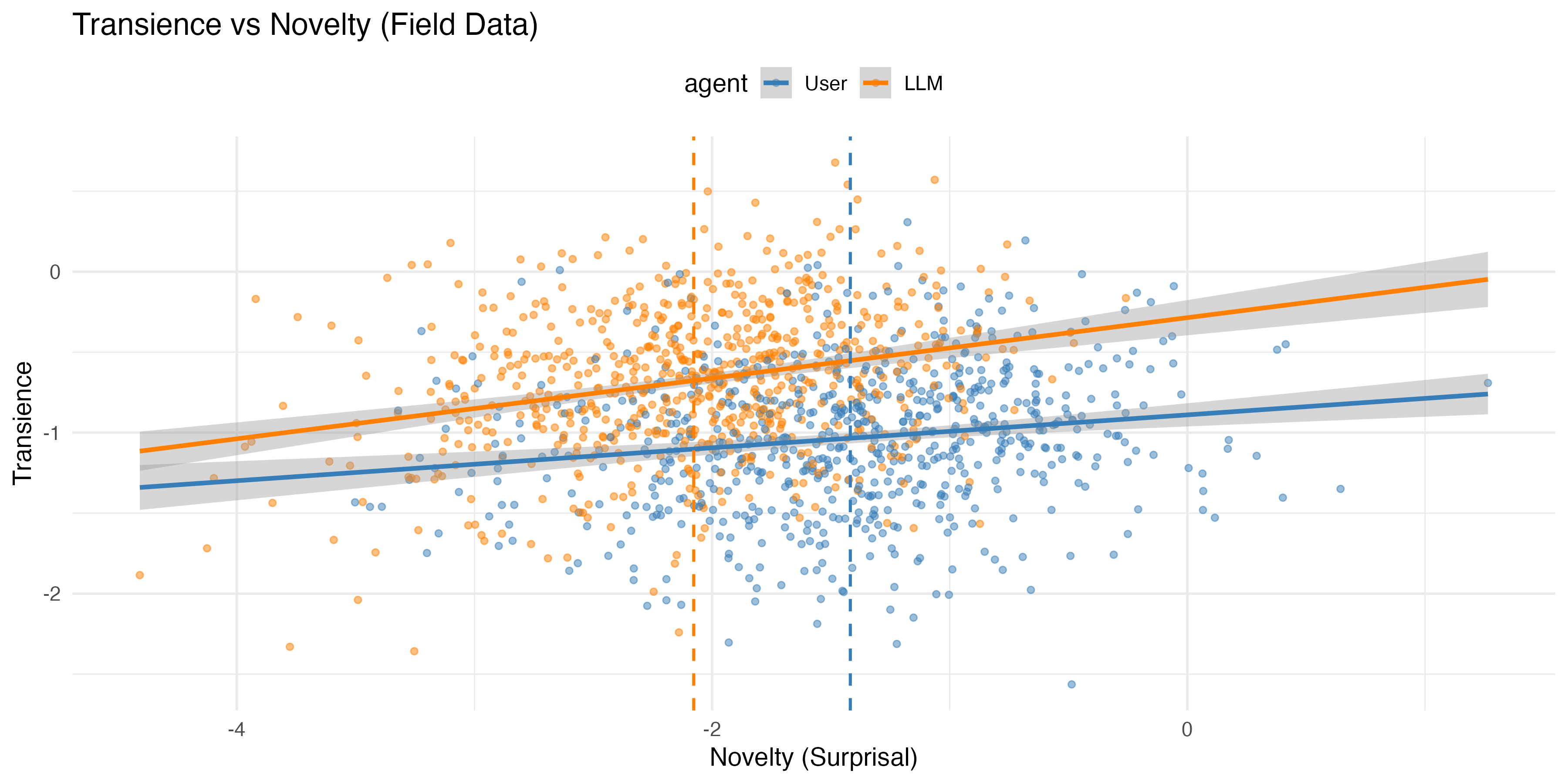}
  \caption{Transience as a function of novelty, with regression lines by agent.}
  \label{fig:transience_vs_novelty}
\end{figure}
\end{document}